\documentclass[10pt]{article}

\usepackage{graphicx}
\usepackage{amsmath}
\usepackage{amsfonts}
\usepackage[titletoc, toc, title]{appendix}
\usepackage{authblk}
\usepackage{geometry}
\usepackage{booktabs} 
\usepackage[linesnumbered,algoruled,boxed,lined]{algorithm2e}
\usepackage{enumitem}
\usepackage{listings}
\usepackage{caption} 
\usepackage[bottom,hang,flushmargin]{footmisc}
\begin{document}

\title{Recurrent Binary Embedding for GPU-Enabled Exhaustive Retrieval from Billion-Scale Semantic Vectors}

\author{Ying Shan}
\author{Jian Jiao}
\author{Jie Zhu}
\author{JC Mao}
\affil{Bing Ads of AI \& Research Group, Microsoft Corp.}

\maketitle
\begin{abstract}
Rapid advances in GPU hardware and multiple areas of Deep Learning  open up a new opportunity for billion-scale information retrieval with exhaustive search.  Building on top of the powerful concept of semantic learning, this paper proposes a Recurrent Binary Embedding (RBE) model that learns compact representations for real-time retrieval.  The model has the unique ability to refine a base binary vector by progressively adding binary residual vectors to meet the desired accuracy.  The refined vector enables efficient implementation of exhaustive similarity computation with bit-wise operations, followed by a near-lossless $k$-NN selection algorithm, also proposed in this paper.  The proposed algorithms are integrated into an end-to-end multi-GPU system that retrieves thousands of top items from over a billion candidates in real-time.

The RBE model and the retrieval system were evaluated with data from a major paid search engine.  When measured against the state-of-the-art model for binary representation and the full precision model for semantic embedding, RBE significantly outperformed the former, and filled in over $80\%$ of the AUC gap in-between.  Experiments comparing with our production retrieval system also demonstrated superior performance.

While the primary focus of this paper is to build RBE based on a particular class of semantic models, generalizing to other types is straightforward, as exemplified by two different models at the end of the paper.
\end{abstract}

\maketitle

\section{Introduction}
\label{Lab:Introduction}
In the age of information explosion, human attention remains a single threaded process.  As the key enabler finding where to focus, information retrieval (IR) becomes ubiquitous, and is at the heart of modern applications including web search, online advertising, product recommendation, digital assistant, and personalized feed.

In IR's almost 100-year history~\cite{sanderson2012history}, a major milestone around $60$s-$70$s~\cite{salton1975vector} was to view queries and documents as high dimensional term vectors, and measure lexical similarity using the cosine coefficient.  Since then, the mainstream development continuously refined the weights of the terms.  Seminal works included \emph{term frequency} ($t\!f$), combined $t\!f$ and inverted document frequency ($id\!f$), the \emph{binary independence model}~\cite{yu1976precision}, and the less probabilistic but highly effective BM25~\cite{robertson2009probabilistic}.

\emph{Latent semantic analysis} (LSA)~\cite{deerwester1990indexing} marked the beginning of matching queries and documents at the semantic level.  As a result, queries and documents relevant in semantics can score high in similarity, even though they are lexically disjoint.  Inspired by LSA, a number of probabilistic topic models were proposed and successfully applied to semantic matching~\cite{hofmann1999probabilistic, blei2003latent, wei2006lda}.

Recent trends have seen the blending of the latent analysis with DNNs.  Models such as \emph{semantic hashing}~\cite{salakhutdinov2007semantic} and \emph{word2vec}~\cite{W2V:13} learned word and phrase embeddings through various DNNs.  Due to weak semantic constraints in the training data, they are not strictly semantic.  However, the connections empowered latent analysis with the latest technologies and tools developed in the DNN community.

The \emph{deep structured semantic model} (DSSM)~\cite{huang2013learning} was among the first DNNs that learned truly semantic embeddings based on search logs.  Applying user clicks as labels enabled a discriminative objective function optimized for differentiating the relevant from the irrelevant.  It significantly outperformed models with objectives only loosely coupled with IR tasks.  DSSM was later upgraded to the \emph{convolutional latent semantic model} (CLSM)~\cite{shen2014latent}, by adding word sequence features through a convolution-pooling structure.

While semantic embedding is advantageous as a representation, online retrieval has to solve the high-dimensional $k$-nearest neighbor ($k$-NN) problem.  The key challenge is to achieve a balanced goal of retrieval performance, speed, and memory requirement, while dealing with the \emph{curse of dimensionality}~\cite{weber1998quantitative, aggarwal2001surprising}.

This paper proposes a novel semantic embedding model called \emph{Recurrent Binary Embedding} (RBE), which is designed to meet the above challenge.  It is built on top of CLSM, and inherits the benefits of being discriminative and order sensitive.  The representation is compact enough to fit over a billion documents into the memory of a few consumer grade GPUs.  The similarity computation of RBE vectors can fully utilize the SIMT parallelism, to the extent that a $k$-NN selection algorithm based on exhaustive search\footnote{Sometimes referred to as \emph{brute-force} search.  They will be referred to interchangeably} is feasible in the range of real-time retrieval.  As a result, the curse of dimensionality that has been haunting the decades-long research of \emph{approximate nearest neighbor} (ANN)~\cite{friedman1977algorithm, datar2004locality} has little effect\footnote{Due to the curse of dimensionality, ANNs exploring a $5\%$ fraction of a $128$-dimensional hyper unit-cube have to search $98\%$ of each coordinate~\cite{wieschollek2016efficient}.  Brute-force search that matches against the entire document set is not subject to this predicament}.


To our best knowledge, this is the first time a brute-force $k$-NN is applied to a billion-scale application, sponsored search in this case, for real-time retrieval.  A salient property of RBE is that the retrieval accuracy can be optimized based on hardware capacity.  Looking ahead, we expect the baseline established in this paper  will be continuously refreshed by more powerful and cheaper hardware, in addition to algorithmic advances.

After presenting details of RBE and the retrieval system, more related work will be reviewed and compared at the end of the paper.

\section{Sponsored Search}
\label{sponsored_search}
RBE is discussed in the context of \emph{sponsored search} of a major search engine.  Readers can refer to \cite{edelman2005internet} for an overview on this subject.  In brief, sponsored search is responsible for showing \emph{ads} (advertisement) alongside organic search results.  There are three major agents in the ecosystem including the user, the advertiser, and the search platform.  The goal of the platform is to display a list of ads  that best match user's intent.  Below is the minimum set of key concepts for the discussions that follow.
\begin{description}[style=unboxed,leftmargin=1.5em]
	\item[Query:] A text string that expresses user intent.  Users type queries into the search box to find relevant information
	\item[Keyword:] A text string that expresses advertiser intent.  Keywords are not visible to users, but play a pivotal role in associating advertiser intent with user intent
	\item[Impression:]  An ad being displayed to a user, on the result page of the search engine
	\item[Click:] An indication that an impressed ad is clicked by a user
\end{description}

On the backend of a paid search engine, the number of keywords are typically at the scale of billions.  IR technologies are applied to reduce the amount of keywords sent to the downstream components, where more complex algorithms are used to finalize the ads to display.  A click event is recorded when an impressed ad is clicked on.

To be consistent with the above context, \emph{keyword} is used instead of \emph{document} throughout the paper.  The query and the keyword associated with a click event is referred to as a \emph{clicked pair}, which is the source of positive samples for many paid search models, including RBE.
\section{Problem Statement}
\label{sec:problemstatement}
Our goal is to find a vector representation that balances the retrieval performance, speed, and storage requirement as mentioned in Sec.~\ref{Lab:Introduction}.  For the subsequent discussions, $q$ and $k$ will be used to denote query and keyword, respectively.   

As mentioned in Sec.~\ref{Lab:Introduction}, there are many ways of representing query and keywords with vectors. This paper primarily focuses on semantic vectors produced by the CLSM model~\cite{shen2014latent}.

\subsection{A Brief Recap on CLSM}
\begin{figure}
	\centering
	\includegraphics[scale=0.8]{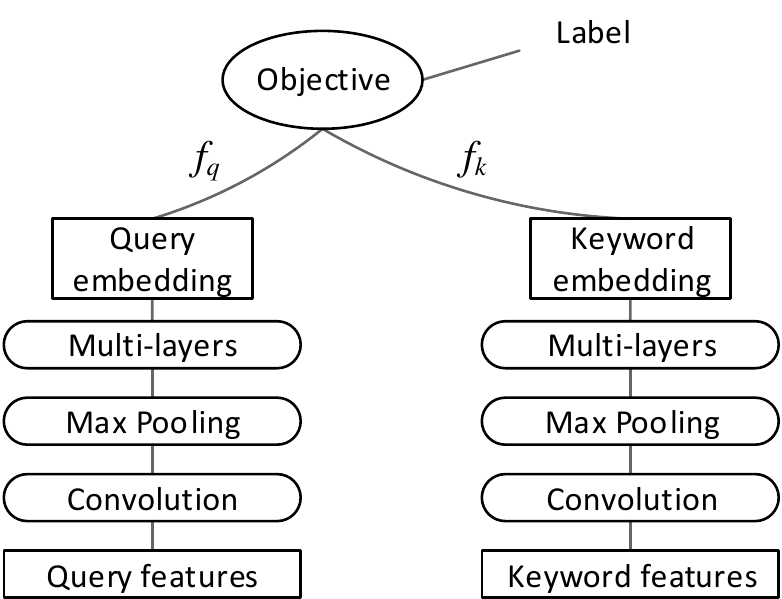}
	\caption{The CLSM model architecture}
	\label{fig:TheCLSMModelArchitecture}
\end{figure}
Fig.~\ref{fig:TheCLSMModelArchitecture} is the high-level architecture of the CLSM model, which consists of two parallel feed-forward networks.  On the left side, query features are mapped from a sparse representation called tri-letter gram\footnote{A form of $3$-shingling at the letter level} to a real-valued vector, through transforms including \emph{convolution}, \emph{max pooling}, and multiple hidden layers (\emph{multi-layers}). The right side for keywords uses the same transformation structure, but with a different set of parameters.

For training, a positive sample is constructed from a clicked pair as mentioned in Sec.~\ref{sponsored_search}.  Negative samples are sometimes generated from positive samples through cross sampling\footnote{Given two pairs of positive samples $\left\langle q_1, k_1\right\rangle$ and $\left\langle q_2, k_2\right\rangle$, cross sampling produces two negative samples $\left\langle q_1, k_2\right\rangle$ and $\left\langle q_2, k_1\right\rangle$}. For each sample $s$, CLSM produces a pair of vectors as in Fig.~\ref{fig:TheCLSMModelArchitecture}, where $\mathbf{f}_q(s)$ and $\mathbf{f}_k(s)$ are for the query, and the keyword, respectively.  

A \emph{sample group} $S$ includes one positive sample and a fixed number of negative samples.  It is the basic unit to evaluate the objective function:
\begin{equation}
\label{equ:objective}
	\mathcal{O}(\mathcal{S}; \Theta) = -\sum_{S\in \mathcal{S}}\log P(S),
\end{equation}
where $\mathcal{S}$ is the set of all sample groups, and $\Theta$ is the set of all CLSM parameters.  In the above equation:
\begin{equation}
\label{equ:persampleprobability}
P(S) \equiv P(s^+ | S) = \frac{\exp(\gamma\; \beta(\mathbf{f}_q(s^+), \mathbf{f}_k(s^+)))}{\sum_{s\in S}\exp(\gamma\; \beta(\mathbf{f}_q(s), \mathbf{f}_k(s)))},
\end{equation}
where $P(s^+ | S)$ is the probability of the only positive query-keyword pair in the sample group, given $S$. The smoothing factor $\gamma$ is set empirically on a held-out data set.  The similarity function is defined as the following:

\begin{equation}
\label{equ:similarity}
	\beta(\mathbf{f}_q, \mathbf{f}_k) \equiv \cos(\mathbf{f}_q, \mathbf{f}_k) = \frac{\mathbf{f}_q \cdot \mathbf{f}_k}{\|\mathbf{f}_q\|\;\|\mathbf{f}_k\|}.
\end{equation}
\subsection{Formulating the Goal}
\label{sec:formulatinggoal}
The goal is to design a model with specially constructed embedding vectors $\mathbf{b}_q^u$ and $\mathbf{b}_k^v$, such that objective function in (\ref{equ:objective}) is minimized:
\begin{equation}
\label{equ:optimization}
	\arg\min_\Theta \mathcal{O}(\mathcal{S}; \Theta).
\end{equation}
Unlike real-valued $\mathbf{f}_q$ and $\mathbf{f}_k$,  $\mathbf{b}_q^u$ and $\mathbf{b}_k^v$ can be decomposed into a series of $u\!+\!1$ and $v\!+\!1$ binary vectors.  The similarity function in (\ref{equ:persampleprobability}) and (\ref{equ:similarity}) becomes: 
\begin{equation}
\label{equ:binarysimilarity}
	\beta(\mathbf{b}_q^u, \mathbf{b}_k^v) \equiv \cos(\mathbf{b}_q^u, \mathbf{b}_k^v).
\end{equation}
The full definition of $\mathbf{b}_q^u$ and $\mathbf{b}_k^v$ is deferred to Sec.~\ref{sec:modelandarchitecture}.  The following is an example to motivate the goal of binary decomposition.  Suppose we have:
\begin{align}
	&\mathbf{b}_q^1 = \mathbf{b}^0_q + \mathbf{d}^0_q \label{equ:motivationalexample}\\ 
	&\mathbf{b}_k^1 = \mathbf{b}^0_k + \mathbf{d}^0_k, \nonumber
\end{align}
where $\mathbf{b}^0_q$, $\mathbf{d}^0_q$, $\mathbf{b}^0_k$, $\mathbf{d}^0_k$ are binary vectors, the cosine similarity becomes:
\begin{equation}
\label{equ:example}
\begin{aligned}
	\cos(\mathbf{b}_q^1, \mathbf{b}_k^1) &= \frac{\mathbf{b}_q^1 \cdot \mathbf{b}_k^1}{\|\mathbf{b}_q^1\|\; \|\mathbf{b}_k^1\|} \\
	&= \frac{(\mathbf{b}^0_q \cdot \mathbf{b}^0_k + \mathbf{b}^0_q \cdot \mathbf{d}^0_k + \mathbf{d}^0_q \cdot \mathbf{b}^0_k + \mathbf{d}^0_q \cdot \mathbf{d}^0_k)}
	{\|\mathbf{b}_q^1\|\; \|\mathbf{b}_k^1\|}.
\end{aligned}
\end{equation}
As demonstrated in (\ref{equ:example}), the binary decomposition turns similarity computation into a series of dot products of binary vectors, which can be implemented efficiently on modern hardware including GPUs.  The hardware enabled computation, together with the compact representation as binary vectors, form the foundation of our work.
\section{Recurrent Binary Embedding}
\label{sec:recurrentbinaryembedding}
To construct the embedding vectors in (\ref{equ:binarysimilarity}), we propose a deep learning model called \emph{Recurrent Binary Embedding}, or RBE.  The model is learned from a training set with clicked pairs.  The learned model is applied to generate query and keyword embedding vectors for retrieval.
\subsection{The Model and the Architecture}
\label{sec:modelandarchitecture}
\begin{figure}
	\centering
		\includegraphics[scale=0.8]{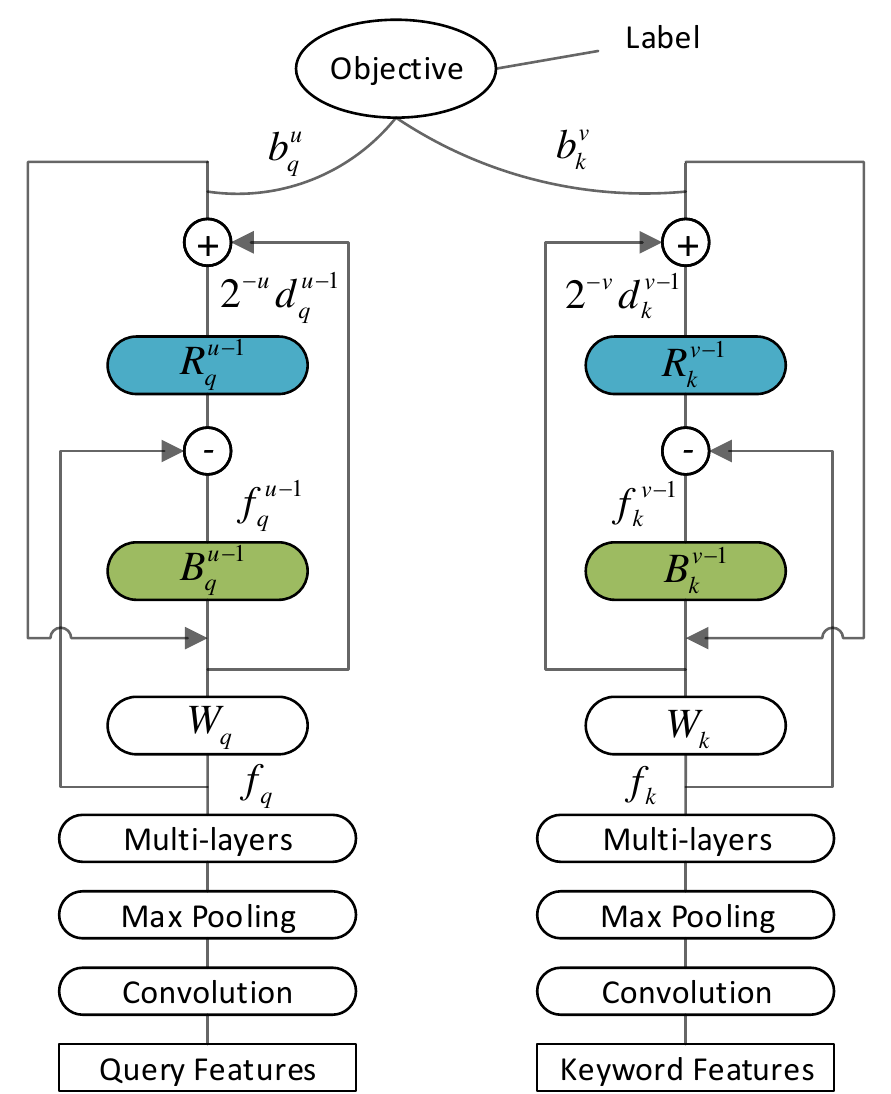}
	\caption{The Recurrent Binary Embedding (RBE) model}
	\label{fig:RBEmodel}
\end{figure}
Fig.~\ref{fig:RBEmodel} is the model architecture of RBE.  As compared with the CLSM model in Fig.~\ref{fig:TheCLSMModelArchitecture}, RBE also has two separate routes for embedding, and shares the same forward processes up to the multi-layers transformations.  The parts beyond the multi-layers will be referred to as \emph{RBE layers} hereafter, and are formulated by the following equations:

\begin{align}
	&\mathbf{b}^0_i = \rho\;(\mathbf{W}_i \cdot \mathbf{f}_i) \label{equ:RBE1}\\
	&\mathbf{f}^{t-1}_i = \tanh(\mathbf{B}^{t-1}_i \cdot \mathbf{b}^{t-1}_i) \label{equ:RBE2}\\
	&\mathbf{d}^{t-1}_i = \rho\;(\mathbf{R}^{t-1}_i \cdot (\mathbf{f}_i - \mathbf{f}^{t-1}_i)) \label{equ:RBE3}\\
	&\mathbf{b}^{t}_i = \mathbf{b}^{t-1}_i + \left(\frac{1}{2}\right)^{t}\mathbf{d}^{t-1}_i \label{equ:RBE},
\end{align}
where $\mathbf{f}_i\in \mathbb{R}^m$, $t\geq 1$ is the time axis discussed later in Sec.~\ref{sec:timeaxis}, and $\mathbf{b}_i^{t-1}\in\{-1,1\}^n$.  Bias terms are dropped for simplicity.  The key idea behind the above equations is to construct the binary decomposition $\mathbf{b}^{t}_i$ by maximizing the information extracted from the real-valued vectors $\mathbf{f}_i$.  A number of intermediate vectors are involved during the training process to achieve this objective.
\begin{description}[style=unboxed,leftmargin=1.5em]
	\item[The base vector] Equation (\ref{equ:RBE1}) is where the real-valued embedding vector $\mathbf{f}_i$ (in float)  is transformed by an $n\times m$ matrix $\mathbf{W}_i$, where the index $i$ is either $q$ or $k$.  It is then mapped into a binary \emph{base vector} $\mathbf{b}^0_i$ through the binarization function $\rho$, discussed with details in Sec.~\ref{sec:binarization}
	\item[The reconstructed vector] Equation (\ref{equ:RBE2}) converts the $n$ dimensional binary vector $\mathbf{b}_i^{t-1}$ back to an $m$ dimensional \emph{reconstructed vector} in float.  The transformation is through an $m\times n$ matrix $\mathbf{B}^{t-1}_i$, followed by an element-wise $\tanh$
	\item[The residual vector] Equation (\ref{equ:RBE3}) transforms the difference between $\mathbf{f}_i$ and $\mathbf{f}^{t-1}_i$ by an $n\times m$ matrix $\mathbf{R}^{t-1}_i$, followed by $\rho$.  The transformed binary vector $\mathbf{d}^{t-1}_i$ is the \emph{residual vector}, because it is transformed from the residual between the original embedding vector and the reconstructed vector
	\item[The RBE embedding] Equation (\ref{equ:RBE}) creates a \emph{refined vector} by recursively adding residual vectors from the previous time stamps, multiplied by a \emph{residual weight} $2^{-t}$ detailed in Sec.~\ref{sec:residualweight}.  The last refined vector is the \emph{RBE embedding}.    The binary vectors adding up to form the RBE embedding are the \emph{ingredient vectors}
\end{description}
At the top of Fig.~\ref{fig:RBEmodel}, RBE embeddings from both sides are used to evaluate the objective function as described in (\ref{equ:optimization}) and (\ref{equ:binarysimilarity}).
\subsection{The Binarization Function}
\label{sec:binarization}
The binarization function $\rho$ in (\ref{equ:RBE1}) and (\ref{equ:RBE3}) plays an important role in training the RBE model.  In the forward computation, it converts float input $x$ into a binary value of either $-1$ or $+1$ as the following:
\begin{equation}
\label{equ:binarizationfunction}
	\rho(x)\equiv \mbox{sign}(x) = 
	\begin{cases}
      -1, & \text{if}\ x \le 0 \\
      1, & \text{otherwise.}
    \end{cases}
\end{equation}
The backward computation is problematic since the gradient of the $\mbox{sign}$ function vanishes almost everywhere. Different gradient estimators were proposed to address the problem.

The straight-through estimator takes the gradient of the identity function as the estimate of $\rho'(x)$~\cite{DBLP:journals/corr/BengioLC13}.  A variant of this estimator, found to have better convergence property, sets $\rho'(x)=1$ when $|x| \le 1$ and $\rho'(x)=0$ otherwise~\cite{courbariaux2016binarized}.  An unbiased gradient estimator was proposed in \cite{DBLP:journals/corr/BengioLC13}, but did not reach the same level of accuracy in practice. 

Another estimator mimics the discontinuous $\mbox{sign}$ function with an annealing $\tanh$ function in backward propagation \cite{courbariaux2016binarized,DBLP:journals/corr/CaoL0Y17}. The annealing $\tanh$ function approaches a step function when the \emph{annealing slope} $\alpha$ increases: 
\begin{equation}
\lim_{\alpha \to \infty} \tanh(\alpha x) = \mbox{sign}(x).
\end{equation}
The slope is initialized with $\alpha=1$, and is increased gently to ensure convergence.  Sec.~\ref{sec:binarylayer} compares the performance of the above estimators referred to later as \emph{straight-through}, \emph{straight-through variant}, and \emph{annealing tanh}, respectively.
\subsection{Residual Weights}
\label{sec:residualweight}
The presence of the residual weight in (\ref{equ:RBE}) may seem natural and intuitive.  However, a closer look reveals a profound implication to the richness of the RBE embedding.

Recall that each dimension of a binary vector in (\ref{equ:RBE}) takes a value of either $-1$ or $1$.  When two binary vectors are added without the residual weight as in (\ref{equ:motivationalexample}), each dimension will end up with a value in $\{-2,0,2\}$.  However, if the same vectors are added with the residual weight, the set of possible values becomes $\{-1.5,-0.5, 0.5, 1.5\}$.  As a result, the cardinality of all RBE embeddings with two ingredient vectors increases from $3^n$ to $4^n$.  In general, it can be proved\footnote{Not elaborated here due to space limit} that for RBE embeddings with $j=t+1$ ingredient vectors, the cardinality grows $(2^j/(j+1))^n$ times by introducing residual weights.  This leads to a substantial boost in accuracy as demonstrated later in Sec.~\ref{sec:witwithoutResidualWeights}.  The base of the weighting schema is set to $\frac{1}{2}$ due to equal distance values for each dimension, and hardware enabled implementation with a bit-wise operation.
\subsection{The Recurrent Pattern}
RBE gets its name from the looped pattern exhibited in the equations of (\ref{equ:RBE2}), (\ref{equ:RBE3}), and (\ref{equ:RBE}).  The pattern is also obvious in Fig.~\ref{fig:RBEmodel}, where $\mathbf{B}^{t-1}_i$ and $\mathbf{R}^{t-1}_i$ alternate to generate the reconstructed vector and the residual vector, and to refine the base vector iteratively.  While the word ``recurrent'' may imply connections with other network structures such as RNN and LSTM, the analogy does not go beyond the looped pattern.

The primary difference is in the purpose of the repeating structures.  For RNN and LSTM-alike models, the goal is to learn a persistence memory unit.  As a result, the transformations share the same set of parameters from $t\!-\!1$ to $t$. In contrast, RBE has the flexibility to decide whether the parameters of $\mathbf{B}^{t-1}_i$ and $\mathbf{R}^{t-1}_i$ should be time varying or fixed.  This helps to optimize the system under various design constraints.
\subsection{Two Time Axes}
\label{sec:timeaxis}
RBE has two time axes $u$ and $v$ for the query side, and the keyword side, respectively.  They are the same as $t$ in the equations of (\ref{equ:RBE2}), (\ref{equ:RBE3}), and (\ref{equ:RBE}).  The two time axes are independent, and can have different numbers of iterations to produce RBE embeddings.  This is another flexibility RBE provides to meet various design constraints.  The benefit will be made clear in Sec.~\ref{sec:effectofdesignchoices}, where RBE models with different configurations are implemented and compared.
\section{RBE-based Information Retrieval}
\label{seq:rbeGIR}
\begin{figure}
	\centering
		\includegraphics[scale=0.8]{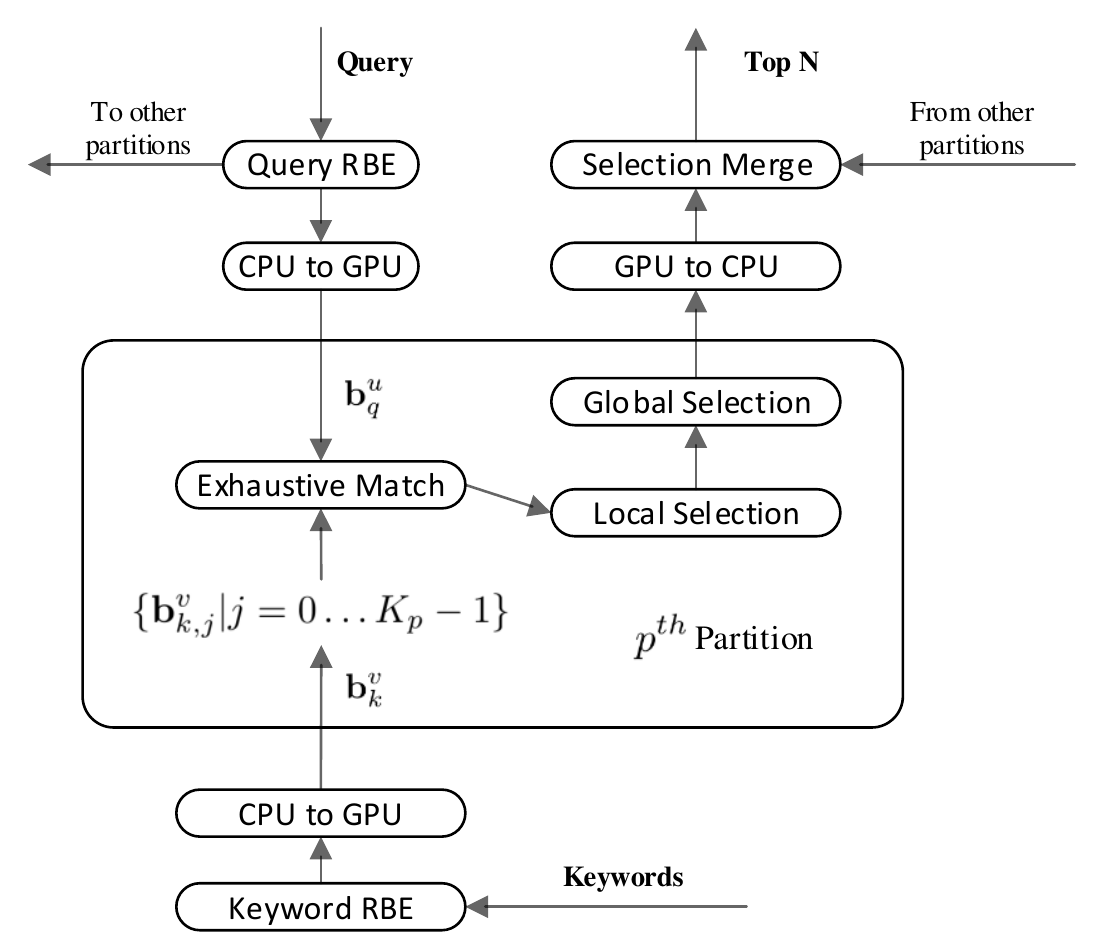}
	\caption{The RBE GPU-enabled Information Retrieval (rbeGIR) system}
	\label{fig:e2esystem}
\end{figure}
A system for keyword retrieval is built based on RBE embeddings.  Fig.~\ref{fig:e2esystem} outlines the high-level architecture of the system, referred to as RBE GPU-enabled Information Retrieval, or rbeGIR.

The system uses multiple GPUs to store and process the RBE embeddings.  The rounded rectangle in the middle of Fig.~\ref{fig:e2esystem} represents the key components of the $p^{th}$ GPU, where the corresponding data partition stores RBE embeddings represented by $\{\mathbf{b}^v_{k,j} | j = 0\ldots K_p-1\}$.  The raw keywords, $K_p$ in total for the $p^{th}$ partition, are transformed offline to vectors through the keyword side of the model in Fig.~\ref{fig:RBEmodel}.  They are uploaded to GPU memory from CPU memory as illustrated on the bottom of Fig.~\ref{fig:e2esystem}.

At run time, a query embedding $\mathbf{b}^u_q$ is generated on-the-fly by the query side of the RBE model.  The same embedding is also sent to other GPUs as shown on the upper left side of the figure.  The \emph{exhaustive match} component inside the GPU is where the similarity function in (\ref{equ:binarysimilarity}) is evaluated for all pairs of $\langle\mathbf{b}^u_q,  \mathbf{b}^v_{k,j}\rangle$.  The similarity values are used to guide the per thread \emph{local selection} and the per GPU \emph{global selection} to find the best keywords from the $p^{th}$ partition.  The results from all GPUs will be used to produce the top $N$ keywords, through the \emph{selection merge} process\footnote{The selection merge process actually relies on a GPU-based Radix sort, which is omitted in Fig.~\ref{fig:e2esystem} for simplicity}.   
\subsection{Dot Product of RBE Embeddings}
\label{sec:RBEDotProduct}
Section~\ref{sec:formulatinggoal} touched upon the dot product of RBE embeddings with a specific example.  More generally, from (\ref{equ:binarysimilarity}), (\ref{equ:RBE1}), and (\ref{equ:RBE}) we have:
\begin{equation}
\begin{aligned}
	\cos(\mathbf{b}_q^u, \mathbf{b}_k^v) 
	&\propto \frac{1}
	{\|\mathbf{b}_k^v\|} \left(\mathbf{b}^0_q \cdot \mathbf{b}^0_k + \sum^{u-1}_{j=0}\sum^{v-1}_{i=0} \left(\frac{1}{2}\right)^{j+i+2}\mathbf{d}_q^j \cdot \mathbf{d}_k^i\right. \\
	&\left.+\sum^{u-1}_{j=0}\left(\frac{1}{2}\right)^{j+1}\mathbf{b}^0_q\cdot\mathbf{d}_k^j
	+\sum^{v-1}_{i=0}\left(\frac{1}{2}\right)^{i+1}\mathbf{b}^0_k\cdot\mathbf{d}_q^i
	\right),
\end{aligned}
\label{equ:rbedotproduct}
\end{equation}
where the magnitude of the query side embedding is dropped because it's the same for all keywords.  Equation (\ref{equ:rbedotproduct}) decomposes the dot product of RBE embeddings into dot products of binary vectors, which can be implemented with bit-wise operations as the following:
\begin{equation}
\label{equ:binaryexecution}
	\mathbf{x} \cdot \mathbf{y} = (popc(\mathbf{x} \wedge \mathbf{y})\gg 1) + n,
\end{equation}
where $\mathbf{x}$ and $\mathbf{y}$ are vectors in $\{-1, 1\}^n$.  On the right side of (\ref{equ:binaryexecution}), $popc$, ``$\wedge$'', and ``$\gg$'' are the \emph{population count}, XOR, and logical right shift operators, respectively.  Multiplying the residual weights also uses the right shift operator, which is executed at most $u+v$ times by carefully ordering the results of binary dot products\footnote{As an illustrative example, computing $\frac{1}{2} a + \frac{1}{4}b$ costs $3$ shifts with ($a\!\gg\! 1) + (b\!\gg\! 2$), but only $2$ shifts with $(a+b\!\gg\! 1)\!\gg\! 1$}.  Since the keyword side magnitude $\|\mathbf{b}_k^v\|$ is usually precomputed and stored with the RBE embeddings, the computation of cosine similarity boils down to a series of binary operations that can be accelerated with (\ref{equ:binaryexecution}), which enables the exhaustive match component in rbeGIR.
\subsection{Asymmetric Design}
\label{sec:asymmetricdesign}
In general, increasing $u$ and $v$ in (\ref{equ:rbedotproduct}) improves accuracy, but comes with the cost of memory and speed. A key observation is that the memory impact of $u$ is negligible, which suggests an \emph{asymmetric design} with $u>v$. This is feasible thanks to the independent time axes mentioned in Sec.~\ref{sec:timeaxis}. However, adding more ingredient vectors on the query side impacts speed.  The trade-off will be studied later with experiments.
\subsection{Key Advantages}
With RBE embedding, storing one billion keywords needs only $14.90$GB memory, instead of $238$GB using float.  This makes in-memory retrieval possible on a few GPUs.  The similarity function is computed exhaustively for all query-keyword pairs with a $k$-NN selection algorithm discussed in Sec.~\ref{sec:gpuimplementation}, which reduces false negatives to almost zero.  Also because of exhaustive matching, there is no need to make implicit or explicit assumptions about data distributions, resulting in consistent retrieval speed and accuracy.  RBE learns application specific representations, and is more accurate than general purpose quantization algorithms.  It is more accurate than other learning-based binarization algorithms due to the recurrent structure.

In addition, rbeGIR mostly uses primitive integer operations available on consumer grade GPUs. As a result, the implementation on low-end GPUs may even outperform high-end ones with higher FLOPS and price tag. It is also straightforward to port the system to different hardware platforms.
\section{Exhaustive $k$-NN Selection with GPU}
\label{sec:gpuimplementation}
At the core of the rbeGIR system is a GPU-based brute-force $k$-NN selection algorithm designed for billion-scale retrieval.
The selection algorithm starts from a local selection process that relies on a $k$-NN kernel outlined in Algorithm~\ref{algo:knn}.  A kernel is a function replicated and launched by parallel threads on the GPU device, each with different input data.  In Algorithm~\ref{algo:knn}, $I$ is the number of keywords to process per thread, $T_b$ is the number of threads per block\footnote{A GPU block is a logical unit containing a number of coordinated threads and certain amount of shared resource}, $x$ is the block id, $y$ is the thread id, and $z$ is the memory offset to the keyword vectors.
\begin{algorithm}
\caption{The $k$-NN Kernel}
\label{algo:knn}
\KwIn{
	The RBE embedding for the query $\mathbf{b}^u_q$ \\
	The RBE embeddings for keywords $\mathbf{b}^v_{k,j\in\Omega_y}$
	}
\KwOut{Priority queue $p$ containing top similarity scores and their indices}
$z \gets x * T_b * I + y$ \;
$p$.clear() \;
\For{$i \gets 1$ \textbf{to} $I$} {
	$s \gets \cos(\mathbf{b}^u_q$, $\mathbf{b}^v_{k,z}$)\;
    $p$.insert($s$, $z$)\;
    $z \gets z + T_b$\;
}
\Return{$p$}\;
\end{algorithm}
The input $\mathbf{b}^v_{k,j\in\Omega_y}$ denotes the RBE embeddings of all keywords $\Omega_y$ processed by the $y^{th}$ thread.  The cosine similarity function is implemented as described in Eq.~\ref{equ:rbedotproduct} and Sec.~\ref{sec:RBEDotProduct}. 

The returned priority lists  are sent to the global selection process, and the merge selection process as mentioned in Sec.~\ref{seq:rbeGIR}.  Both processes leverage the Radix sort method mentioned in~\cite{merrill2010revisiting}, which is one of the fastest GPU-based sorting algorithms.
\subsection{Performance Optimization}
The design of the brute-force $k$-NN takes into consideration several key aspects of the GPU hardware to save time.  First, the global selection process handles only the candidates in the priority lists.  This avoids extensive read and write operations in the global memory.  Second, a number of sequential kernels are fused into a single thread, which takes the full advantage of thread level registers instead of much slower shared memory and global memory. 

The storage of RBE embeddings is also re-arranged to utilize the warp-based memory loading pattern.  Instead of organizing the embeddings by keywords, the base vectors and residual vectors are grouped and stored separately.  In the case of RBE embeddings with two ingredient vectors, the base vectors are stored first in a continuous memory block, followed by the residual vectors.  This makes it possible for a warp ($32$ consecutive threads) to load $128$ bytes of memory in a single clock cycle, which improves the kernel speed significantly.
\subsection{Negligible Miss Rate}
The $k$-NN kernel computes similarity scores exhaustively for all keywords.  However, the local selection process relies on a priority queue which is a lossy process.  The key insight is that the miss rate of the algorithm is negligible when $N\ll C$, where $N$ is usually in the range of thousands, and the number of candidates $C$ is in the billions.  As explained in Appendix~\ref{sec:appendixSelectionProbability}, for $N=1000$, $C=10^9$, and $I=256$, the probability of missing more than two relevant keywords is less than $0.03\%$.  In practice, setting the length of the priority queue to be $1$ is sufficient.
\section{Experiment Settings}
\label{sec:experimentresults}
\begin{figure}[b]
	\centering
		\includegraphics[scale=0.8]{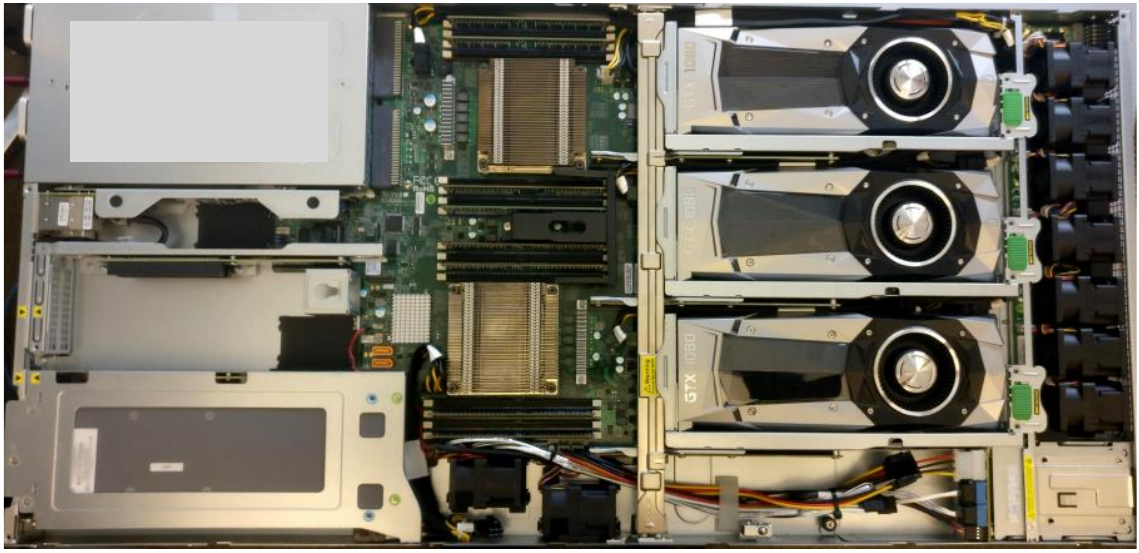}
	\caption{The rbeGIR server with four NVIDIA GeForce GTX$1080$ GPUs, two $8$-core CPUs, and $128$GB DDR memory}
	\label{fig:RBEServer}
\end{figure}
The RBE model was implemented with BrainScript in CNTK\footnote{https://docs.microsoft.com/en-us/cognitive-toolkit/. The BrainScript of RBE will be available as open source in a couple months}, and trained on a GPU cluster.  The CLSM components and the objective function were built from existing CNTK nodes.  The binarization function $\rho$ was implemented as a customized node in CNTK, and exposed to BrainScript.  The recurrent embedding was unfolded into a series of feed-forward layers.  The rbeGIR system was implemented from the ground up on a customized multi-GPU server in Fig.~\ref{fig:RBEServer}.

The convolution layer of the RBE model mapped a sliding group of three words (from either query or keyword input) to a $288$ dimensional (dim) float vector.  Each group of input was a sparse vector of $49292$-dim tri-letter gram.  The max-pooling layer produced an $m=288$-dim vector $\mathbf{f}_i$, which was transformed to an $n=64$-dim base vector $\mathbf{b}^0_i$ as in (\ref{equ:RBE1}).  Time varying matrices of $\mathbf{B}^{t-1}_i$ and $\mathbf{R}^{t-1}_i$ in (\ref{equ:RBE2}) and (\ref{equ:RBE3}) were $288\times 64$, and $64\times 288$, respectively.  Multi-layers in Fig.~\ref{fig:RBEmodel} were not used due to limited performance gain. 

The rbeGIR system used $256$ threads in a single GPU block, where each thread launched the $k$-NN kernel to process $I=256$ keywords.  The total number of blocks was $136\times 136$, arranged on a two dimensional grid.

The experiments were based on the data collected from our paid search engine. The training data for the RBE model contained $175$ million (M) unique clicked pairs sampled from two-year's worth of search logs. Adding $10$ negative samples generated per clicked pair through cross sampling, the total number of training samples amounted to $1925$M.  The validation data had $1.2$M pairs sampled a month after to avoid overlap. The test data consisted of $0.8$M pairs labeled by human judges. Pairs labeled\footnote{See Sec.~\ref{sec:qualityofretrieval} for the details of labels} as \emph{bad} were considered as negative samples, and the rest were considered as positive ones.
\section{Main Results}
\label{sec:mainresults}
The main results are reported based on a $64$-dim RBE model of our choice, referred to hereafter as rbe*.  The model used three ingredient vectors for queries, and two for keywords.  Since ingredient vectors are stored separately, each dimension of the RBE embedding used three bits for queries, and two for keywords.  Straight-through variant and residual weights were applied across all models mentioned in this section.
\subsection{Model Accuracy}
Four models in Table~\ref{tab:accuracymodels} were evaluated against rbe* with accuracy.  The first model, referred to as m\_1, is a CLSM model with $64$-dim embedding layers in float\footnote{Specifically, the CLSM model has a $64$-dim $\tanh$ layer on top of the max pooling layer as described in Sec.~\ref{sec:experimentresults}}.  This sets the upper bound for binarization models including RBE.  The m\_2 model replaces the embedding layers in m\_1 with binarization layers using the state-of-the-art straight-through variant.  The m\_3 model is the same as m\_2 but with $128$-dim embedding, which represents the performance of $2$-bit binarization without changing the structure.  The m\_5 model has full precision embedding for queries, and RBE embedding with two ingredients for keywords.  It sets the upper bound for RBE models with $2$-bit binarization like rbe*.
\begin{table}[h]
\caption{Models configurations for accuracy comparison} 
\centering 
\begin{tabular}{l c c c} 
\hline 
\multicolumn{1}{c}{Model} & Dimension & $q$ (bits) &  $k$ (bits) \\ 
\hline 
m\_1: full precision CLSM  & 64 & 32 &  32 \\
m\_2: state-of-the-art & 64 & 1 & 1\\
m\_3: state-of-the-art & 128 & 1 & 1 \\ 
m\_4: rbe* & 64 & 3 & 2 \\
m\_5: hybrid rbe & 64 & 32 & 2 \\
\hline
\end{tabular}
\label{tab:accuracymodels} 
\end{table}

Two metrics are used for comparison in Table~\ref{tab:pf170m}.  The first metric is \emph{log loss} defined in (\ref{equ:objective}), which measures the difference between the distributions of the predicted similarity and the click labels.  Log loss is applied to both the training set and the validation set.  The second metric is ROC AUC measured on the test data set with human labels.  The last column of the table is the \emph{AUC lift} defined based on the AUC difference $0.0198$ between m\_3 and m\_1, referred hereafter as the \emph{reference gap}.  As an example, the lift for rbe* is $0.0159/0.0198*100=80.30\%$, where the numerator is the AUC improvement from m\_3 to rbe*.
\begin{table}[h]
\caption{Accuracy of embedding models} 
\centering 
\begin{tabular}{l c c c} 
\hline 
\multicolumn{1}{c}{Model} & Log Loss & ROC AUC & AUC Lift \%\\
\hline
m\_1: full precision CLSM  & 0.0293 & 0.8044 & -\\
m\_2: state-of-the-art & 0.0481 & 0.7719 & -64.14\\
m\_3: state-of-the-art  & 0.0425 &  0.7846  & -\\ 
m\_4: rbe*  & 0.0312 & 0.8005 & 80.30\\ 
m\_5: hybrid rbe  & 0.0311 & 0.8011 & 83.33\\
\hline 
\end{tabular}
\label{tab:pf170m} 
\end{table}

Observing from Table~\ref{tab:pf170m}, the 1-bit increase per dimension improves the accuracy significantly from m\_2 to m\_3.  Without increasing bit per dimension for keywords, rbe* lifts the AUC by more than the amount from m\_2 to m\_3, and is only $3.03\%$ away from the hybrid upper bound of m\_5.   The log loss values exhibit similar gains.
\subsection{The Quality of Retrieval}
\label{sec:qualityofretrieval}
RbeGIR was evaluated against our production retrieval system\footnote{Unfortunately, the details of the system is not available for publishing} with the quality of returned keywords. The production setting included one with the same amount of memory (prod\_1), and another with the same amount of keywords (prod\_2). Since rbeGIR does not use extra indexing structure, only the amount of memory used for keyword vectors were counted for prod\_1. The rbeGIR system stored the embeddings of $1.2$ billion unique keywords.
\begin{table}[h]
\caption{Top five results by quality} 
\centering 
\begin{tabular}{l c c c c} 
\hline 
Baseline & Bad & Fair & Good & Excellent \\
\hline 
prod\_1 & -52.37 & -9.73 & 18.52 & 18.83\\ 
\hline 
prod\_2 & -35.26 & 3.32 & 11.19 & 4.39\\ 
\hline
\end{tabular}
\label{tab:labelresults} 
\end{table}

Table~\ref{tab:labelresults} reports the average quality of the top five keywords returned from each of $2000$ queries.  Based on a production quality guideline, query-keyword pairs were manually judged with a score of \emph{bad}, \emph{fair}, \emph{good}, or \emph{excellent}.  Each column in the table is the percentage change between the counts of query-keyword pairs from rbeGIR and the baseline system by scores.  As an example, rbeGIR retrieved $18.52\%$ more good pairs than prod\_1.  From Table~\ref{tab:labelresults}, rbeGIR has significantly reduced the amount of bad pairs for both prod\_1 and prod\_2.  It found less fair pairs than prod\_1, but otherwise substantially more good and excellent pairs than the production baselines.  It was also observed that there were about $8\!-\!11\%$ overlap of the good or excellent pairs between rbeGIR and the baselines.
\subsection{Recall and Latency}
To evaluate the recall, 10000 queries were first matched offline with $1.2$ billion keywords through exact nearest neighbor, using RBE embeddings generated by the rbe* model.  The per query recall @1000 is defined by the total number of top keywords overlapping with the relevant keywords, divided by $1000$.  It was observed that the average recall @1000 for rbeGIR is $99.99\%$, which is expected per Appendix~\ref{sec:appendixSelectionProbability}.

The latency for rbe\_2 and rbe* are on average $29.92$ms, and $31.17$ms, respectively.  Both models are in the range of real-time retrieval.  Adding one more bit from rbe\_2 to rbe* on the query side increased the query time by $4.18\%$.  The latency of the full precision model was measured on a down-sampled keyword set ($20$M) due to the memory limit.  As compared with the rbe* model using the same keyword set, the latency of the full precision model was around ten times higher.
\section{Effects of Design Choices}
RBE and the rbeGIR system require a limited amount of tuning.  Only a few design choices were experimented before finalizing the rbe* model in Sec.~\ref{sec:mainresults}.  
\label{sec:effectofdesignchoices}
\subsection{Number of Bits per Dimension}
Table~\ref{tab:rbe_iter_effect} lists models similar to rbe* but with different number of bits per dimension.  These models were created by adjusting the number of iterations $u$ and $v$.  The m\_2 model is the same as in Table~\ref{tab:accuracymodels}, with one bit for $q$ and $k$.  As shown in Table~\ref{tab:rbe_iter_effect}, model accuracy improves as the number of bits per dimension increases.
\begin{table}[h]
\centering 
\caption{Model accuracy with varying bits per dimension} 
\begin{tabular}{l l l | c c c} 
\hline 
\multicolumn{3}{c}{Model} & Log Loss & ROC AUC \\
\hline 
m\_2: &$q=1$ &$k=1$  & 0.0481 & 0.7719 \\
rbe\_2: &$q=2$ &$k=2$  & 0.0333 & 0.7972 \\ 
rbe*: &$q=3$ &$k=2$    & 0.0312 & 0.8005 \\ 
rbe\_3: &$q=3$ &$k=3$   & 0.0294 & 0.8034 \\ 
\hline 
\end{tabular}
\label{tab:rbe_iter_effect} 
\end{table}

The best model rbe\_3 in the table is only $5.05\%$ reference gap away from the m\_1 model in Table~\ref{tab:accuracymodels}.  However, rbe* was chosen over rbe\_3 because adding one more ingredient vector for keywords requires $50\%$ more memory. Comparing rbe* with rbe\_2 demonstrates the advantage of using asymmetric design mentioned in Sec.~\ref{sec:asymmetricdesign}.  Using one more bit on the query side, rbe* gains $16.67\%$ over rbe\_2 without additional memory.
\subsection{Type of Binarization Functions}
\label{sec:binarylayer}
Three binarization algorithms mentioned in Sec.~\ref{sec:binarization} were experimented with for RBE training.  The original straight-through had difficulty converging when $u$ or $v$ was larger than $2$, but the straight-through variant mentioned in Sec.~\ref{sec:binarization} converged consistently.  The results in Table~\ref{tab:binarization_effect} compare the AUC performance of the straight-through variant and annealing tanh.
\begin{table}[h]
\caption{Performance of different binarization methods} 
\centering 
\begin{tabular}{c l l | c c c} 
\hline 
\multicolumn{3}{c}{Model} & ST Variant & Annealing Tanh \\
\hline 
rbe\_1: &$u=0$ &$v=0$  & 0.7719 & 0.7730 \\ 
rbe\_2: &$u=1$ &$v=1$  & 0.7972 & 0.7950 \\
rbe*: &$u=2$ &$v=1$  & 0.8005 & 0.8007 \\
rbe\_3: &$u=2$ &$v=2$   & 0.8034 & 0.8014 \\
\hline 
\end{tabular}
\label{tab:binarization_effect} 
\end{table}

Based on Table~\ref{tab:binarization_effect}, annealing tanh performs better than the straight-through variant on the rbe\_1 model.  However, as the number of iterations increases, the straight-through variant (referred to as ``ST variant'' in the table) shows better performance overall\footnote{The rbe* model is an exception}.  This is likely caused by the small gradient of annealing tanh, especially as the annealing slope increases over time.  With more iterations, the gradient vanishes more easily due to the chain effect, making it hard to improve the binarization layers.

Based on the above experiments, rbe* adopted the straight-through variant for binarization.  
\subsection{Inclusion of Residual Weights}
\label{sec:witwithoutResidualWeights}
Table~\ref{tab:residualweight_effect} reports the AUC performance for three models with and without residual weights.  Using the reference gap defined in Sec.~\ref{sec:mainresults}, the lift in accuracy ranges from $33.33\%$ to $43.43\%$.  Combining Table~\ref{tab:pf170m} and Table~\ref{tab:residualweight_effect}, it can be seen that almost half ($39.89\%$) of the total gain ($80.30\%$) from m\_3 to rbe* is due to residual weights.
\begin{table}[h]
\caption{ROC AUC with and without residual weights} 
\centering 
\begin{tabular}{c l l | c c c c} 
\hline 
\multicolumn{3}{c}{Model} & No Weights & Weights & AUC Lift \%\\
\hline 
rbe\_2: &$q=2$ &$k=2$  & 0.7886 & 0.7972 & 43.43 \\
rbe*: &$q=3$ &$k=2$  & 0.7926 & 0.8005 & 39.89\\
rbe\_3: &$q=3$ &$k=3$   & 0.7945 & 0.8034 & 33.33\\
\hline 
\end{tabular}
\label{tab:residualweight_effect} 
\end{table}
\section{Related Work}
\label{sec:relatedwork}
RBE has binarization layers similar to binary DNNs in~\cite{salakhutdinov2007semantic, courbariaux2016binarized, alemdar2017ternary, rastegari2016xnor}, which  focused on finding optimal ways of binarization through better gradient propagation~\cite{courbariaux2016binarized}, or reformulating the objective~\cite{DBLP:journals/corr/abs-1710-11573}. While some binary DNNS reported performance parity with the full precision counterparts, the gap was substantial in our experiments without RBE.  The difference was probably in the size of the training data.  With nearly two billion training samples, the side effect of binarization as a regularization process was no longer effective, and the gap had to be filled in with additional performance drivers such as RBE, which optimized the model structure.

The $k$-NN selection algorithm is related to a class of ANN algorithms such as KD-tree~\cite{friedman1977algorithm}, FLANN package~\cite{LSHDavid:2014}, \emph{neighborhood graph search}~\cite{NGS:2012}, and \emph{locality sensitive hashing}~\cite{LSH:2004}.  Unlike those algorithms, the exhaustive search is not subject to the curse of dimensionality.  As compared with other brute-force $k$-NNs that have the same property,  it handles billion-scale keywords, while existing methods such as ~\cite{li2015brute, tang2015efficient} mostly dealt with data size in the millions or smaller.

The rbeGIR system was implemented on GPUs. Based on a recent ANN called \emph{product quantization} (PQ)~\cite{jegou2011product}, a billion-scale retrieval system on GPU was proposed in~\cite{wieschollek2016efficient}, and extended later in~\cite{johnson2017billion}. In order to achieve speed and memory efficiency, PQ-based approaches had to drastically sacrifice the resolution of the codebook, and rely on lossy indexing structures. In contrast, the rbeGIR system relies on a near lossless $k$-NN, and a compact representation with high and easy to control accuracy.  It also does not involve extra indexing structures that may require extensive memory.

RBE is related to the efforts such as Deep Embedding Forest~\cite{zhu2017deep} to speed up online serving of DNNs like Deep Crossing~\cite{ying2016DeepCrossing}.  However, the focus there was on simplifying the deep layers, rather than a compact representation of the embedding layer. 

Finally, RBE is remotely related to \emph{residual nets}~\cite{he2016deep}, where the ``shortcuts'' in those models were constructed differently and for different purposes. 
\section{Conclusion and Future Work}
\label{sec:conclusionfuturework}
The RBE model proposed in this paper generates compact semantic representation that can be efficiently stored and processed on GPUs to enable billion-scale retrieval in real-time.  Integrating the RBE representation with a GPU-based exhaustive $k$-NN search, the rbeGIR system is expected to set an early example for IR in the era of powerful GPUs and advanced Deep Learning algorithms.

Being able to learn the RBE representation benefits from the advance of Deep Learning, while being able to process RBE representations in real-time benefits from the advance of GPU hardware.  Together, brute-force IR at billion-scale is within the reach.  What is more interesting in this new era is the paradigm shift in designing IR algorithms.  To tame the curse of dimensionality, the answer may lie in something more straightforward, but better utilizing the ever growing power of hardware.

\begin{figure}
	\centering
		\includegraphics[scale=0.85]{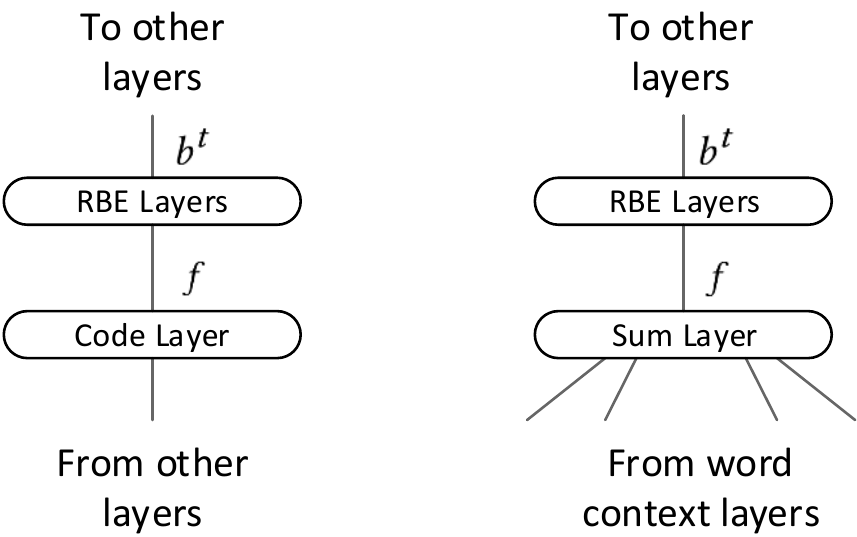}
	\caption{The concept of RBE can be generalized to other networks such as semantic hashing (left) and word2vec (right)}  
	\label{fig:RBEfuture}
\end{figure}

To make the presentation pragmatic and intuitive, RBE is introduced in the context of CLSM and sponsored search.  We conclude this paper by claiming that RBE is not constrained by specific embedding models, and its application is broader.  Part of our future work is to apply the concept of RBE on different network structures such as semantic hashing and word2vec, as illustrated in Fig.~\ref{fig:RBEfuture}, where the RBE layers refer to the layers between $\mathbf{f}_i$ to $\mathbf{b}^t_i$ in Fig.~\ref{fig:RBEmodel}, the \emph{code layer} is the same as in Fig.~$2$ in~\cite{salakhutdinov2007semantic}, and the \emph{sum layer} is the same as in the \emph{CBOW} network of Fig.~$1$ in \cite{mikolov2013efficient}. 
\section{Acknowledgments}
\label{sec:acknowledgements}
The authors would like to thank Ray Lin and Yi Zhang from Bing Ads, and Xiaodong He from MSR for their support and discussions that benefited the development of RBE.
\begin{appendices}
\section{Probability of Miss}
\label{sec:appendixSelectionProbability}
Below is a list of definitions\footnote{Some of the definitions are given before, but are repeated here to avoid cross referencing} used to calculate the probability of missing relevant keywords for the $k$-NN selection algorithm introduced in Sec.~\ref{sec:gpuimplementation}:
\begin{itemize}
\item C -- number of candidates
\item N -- number of relevant keywords, the same as in ``top $N$''
\item I -- number of keywords per thread
\item T -- number of threads, and $T=C/I$
\item M -- number of threads with at least one relevant keyword
\item L -- number of missed relevant keywords
\end{itemize}
The event of missing keywords is defined under the condition of using a priority queue with length equal to $1$.  We start by noticing $L=N-M$, and there are $\Lambda$ combinations of distributing $N$ relevant keywords to $M$ threads\footnote{This is derived using the Stars and Bars method, where the stars are the relevant keywords, and the bars are the threads.}, where $\Lambda=\binom{N-1}{M-1}$.

Suppose that $n_{i,j}$ is the number of relevant keywords in the $j^{th}$ thread of the $i^{th}$ combination. By definition, we have $n_{i,j}\geq 1$ and $\sum_{1\leq j\leq M}n_{i,j}=N$. The number of combinations of having $n_{i,j}$ relevant keywords in the thread with a total of $I$ keywords is $\binom{I}{n_{i,j}}$\footnote{The combinations of $n_{i,j} > I$ are naturally excluded because $\binom{I}{n_{i,j}}=0$}. Since there are $M$ independent threads, the number of combinations of all threads becomes $\prod^M_{j=1}{\binom{I}{n_{i,j}}}$.  Summing up $\Lambda$ mutually exclusive thread combinations leads to:
\begin{equation*}
\sum^{\Lambda-1}_{i=0}\prod^M_{j=1}{\dbinom{I}{n_{i,j}}}.
\end{equation*}
Multiplying by the choices of selecting $M$ threads from the total of $T$ threads, and divided by the number of combinations of selecting $N$ relevant keywords from the entire set of $C$ candidates, the probability of having $M$ threads with at least one relevant keywords is:
\begin{equation}
\label{equ:probabilityofPM}
P(M)=\frac{\dbinom{T}{M}\sum^{\Lambda-1}_{i=0}\prod^M_{j=1}{\dbinom{I}{n_{i,j}}}}{\dbinom{C}{N}}.
\end{equation}
Since $L=N-M$, this is equivalent to the probability of missing $L$ relevant keywords. Table~\ref{tab:probabilityofPL} summarizes the results for $N=1000$, $C=10^9$, and $I=256$ based on (\ref{equ:probabilityofPM}).
\begin{table}[h]
\caption{Probability of missing at most $l$ keywords} 
\centering 
\begin{tabular}{c c c c} 
\hline 
$l$ & 0 & 1 & 2 \\
\hline 
$P(L\leq l)$ & 88.039 & 99.256 & 99.969 \\
\hline 
\end{tabular}
\label{tab:probabilityofPL}
\end{table}

Note that $n_{i,j}$ has to be enumerated in order to use (\ref{equ:probabilityofPM}).  This becomes intractable when the number of missing keywords increases. However, since $P(L \leq 2)$ is already high enough, it is of no interest to go after solutions with higher $l$.
\end{appendices}

\bibliographystyle{acm}
\bibliography{RecurrentBinaryEmbedding}

\end{document}